# PHIBSS: molecular gas, extinction, star formation and kinematics in the z=1.5 star forming galaxy EGS13011166[1]

R.Genzel[1,2,3], L.J.Tacconi[1], J.Kurk[1], S.Wuyts[1], F.Combes[4], J.Freundlich[4], A.Bolatto[5], M.C.Cooper[6], R.Neri[7], R.Nordon[8], F.Bournaud[9], A.Burkert[10,11], J.Comerford[12], P.Cox[2], M.Davis[3], N.M. Förster Schreiber[1], S.García-Burillo[13], J.Gracia-Carpio[1], D.Lutz[1], T.Naab[14], S.Newman[3], A. Saintonge[1], K. Shapiro Griffin[15], A.Shapley[16], A.Sternberg[8] & B.Weiner[17]

[1] *Max-Planck-Institut für extraterrestrische Physik (MPE), Giessenbachstr., 85748 Garching, Germany ( linda@mpe.mpg.de, genzel@mpe.mpg.de )*

[2] *Dept. of Physics, Le Conte Hall, University of California, 94720 Berkeley, USA*

[3] *Dept. of Astronomy, Campbell Hall, University of California, Berkeley, CA 94720, USA*

[4]*Observatoire de Paris, LERMA,* CNRS, *61 Av. de l'Observatoire, F-75014 Paris, France*

[5] *Dept. of Astronomy, University of Maryland, College Park, MD 20742-2421, USA*

[6] *Dept. of Physics & Astronomy, Frederick Reines Hall, University of California, Irvine, CA 92697*

[7]*IRAM, 300 Rue de la Piscine, 38406 St.Martin d'Heres, Grenoble, France*

[8] *Sackler School of Physics and Astronomy, Tel Aviv University, Tel Aviv 69978, Israel*

[9] *Service d'Astrophysique, DAPNIA, CEA/Saclay, F-91191 Gif-sur-Yvette Cedex, France*

[10] *Universitätssternwarte der Ludwig-Maximiliansuniversität , Scheinerstr. 1, D-81679 München, Germany*

[11] *MPG-Fellow at MPE*

---

[1] *Based on observations with the Plateau de Bure millimetre interferometer, operated by the Institute for Radio Astronomy in the Millimetre Range (IRAM), which is funded by a partnership of INSU/CNRS (France), MPG (Germany) and IGN (Spain). Based also on data acquired with the Large Binocular Telescope (LBT). The LBT is an international collaboration among institutions in Germany, Italy, and the United States. LBT Corporation partners are LBT Beteiligungsgesellschaft, Germany, representing the Max- Planck Society, the Astrophysical Institute Potsdam, and Heidelberg University; Istituto Nazionale di Astrofisica, Italy; The University of Arizona on behalf of the Arizona university system; The Ohio State University, and The Research Corporation, on behalf of the University of Notre Dame, University of Minnesota, and University of Virginia.*

[12]*Department of Astronomy & McDonald Observatory, 1 University Station, C1402 Austin, Texas 78712-0259, USA*

[13] *Observatorio Astronómico Nacional-OAN, Observatorio de Madrid, Alfonso XII, 3, 28014 - Madrid, Spain*

[14] *Max-Planck Institut für Astrophysik, Karl Schwarzschildstrasse 1, D-85748 Garching, Germany*

[15] *Space Sciences Research Group, Northrop Grumman Aerospace Systems, Redondo Beach, CA 90278, USA*

[16] *Department of Physics & Astronomy, University of California, Los Angeles, CA 90095-1547, USA*

[17] *Steward Observatory, 933 N. Cherry Ave., University of Arizona, Tucson AZ 85721-0065, USA*

# Abstract

We report matched resolution, imaging spectroscopy of the CO 3-2 line (with the IRAM Plateau de Bure millimeter interferometer) and of the Hα line (with LUCI at the Large Binocular Telescope [LBT]) in the massive z=1.53 main-sequence galaxy EGS13011166, as part of the "Plateau de Bure high-z, blue-sequence survey" (PHIBSS: Tacconi et al. 2013). We combine these data with HST V-I-J-H-band maps to derive spatially resolved distributions of stellar surface density, star formation rate, molecular gas surface density, optical extinction and gas kinematics. The spatial distribution and kinematics of the ionized and molecular gas are remarkably similar and are well modeled by a turbulent, globally Toomre unstable, rotating disk. The stellar surface density distribution is smoother than the clumpy rest-frame UV/optical light distribution, and peaks in an obscured, star forming massive bulge near the dynamical center. The molecular gas surface density and the effective optical screen extinction track each other and are well modeled by a 'mixed' extinction model. The inferred slope of the spatially resolved molecular gas to star formation rate relation, $N = d\log\Sigma_{star\ form} / d\log\Sigma_{molgas}$, depends strongly on the adopted extinction model, and can vary from 0.8 to 1.7. For the preferred mixed dust-gas model we find N=1.14±0.1.

# 1. Introduction

In our Milky Way and nearby external galaxies, stars form in dense, molecular gas clouds (e.g. McKee & Ostriker 2007). In massive disk galaxies averaged over scales of several kiloparsecs or more, the surface densities of cold gas and star formation are empirically well correlated through a simple power-law relation, the 'Kennicutt-Schmidt' (KS) relation (Schmidt 1959, Kennicutt 1998a (K98), Kennicutt & Evans 2012),

$$< \dot{\Sigma}_{starform} > \; = \; < \xi \, \Sigma_{gas}{}^{N} > \quad (1),$$

where N=1.4 (±0.15) if $\Sigma_{gas}=\Sigma_{HI+H2}$ (Kennicutt 1998a,b, 2007), above a threshold of ~10 $M_{\odot}pc^{-2}$. Here the brackets denote averages over the galaxy, beam or region considered. A simple physical motivation for equation (1) comes from considering the star formation rate per volume and making the plausible Ansatz for the volumetric gas to star formation rate relation

$$\dot{\rho}_{*} = \varepsilon_{ff} \frac{\rho_{gas}}{\tau_{ff}} \propto \rho_{gas}^{1.5} \text{ since } \tau_{\rm ff} = a \rho_{gas}^{-1/2} \quad (2),$$

where $\dot{\rho}_{*}, \rho_{gas}$ and $\tau_{ff}$ are the star formation rate per volume, the gas density and the free fall time scale in that volume, and $\varepsilon_{ff}$ is the star formation efficiency per free fall time scale ($\varepsilon_{ff}$ ~0.02, K98, Kennicutt & Evans 2012, Krumholz & Tan 2007, Krumholz, Dekel & McKee 2012). For a constant gas/star formation scale height $h_z$ equation (1) follows then from equation (2), with $< \xi > \; = \; < \varepsilon_{ff} / (a h_z^{1/2}) >$.

From spatially resolved mapping of molecular gas (as traced in the CO 2-1 line in the IRAM HERACLES survey), along with HI (from the VLA THINGS survey), and star formation (from GALEX FUV + SPITZER 24µm continua, or Hα + 24µm continuum) in ~30 nearby disk galaxies Bigiel et al. (2008, 2011), Leroy et al. (2008, 2013) and Schruba et al. (2011) have found that the star formation surface density is correlated most strongly with the molecular column density, rather than with the sum of atomic and molecular column densities. These authors find that the ‘molecular’ KS relation simplifies to a linear relation (N=1±0.15), such that equation (1) becomes $<\Sigma_{starform}> = <\Sigma_{molgas}/\tau_{depl}>$, or $SFR=M_{molgas}/\tau_{depl}$, with a molecular gas depletion time[2] of ~1-2 Gyr (see also Rahman et al. 2012). The residual scatter of the KS-relation is ±0.3 dex (K98, Bigiel et al. 2008).

Saintonge et al. (2011a,b, 2012) have measured the molecular gas content from galaxy integrated CO 1-0 observations in a purely stellar mass selected sample drawn from SDSS ($\log M_*>10$, all masses in this paper are in units of $M_\odot$), including both blue- and red-sequence galaxies. Saintonge et al. confirm the ~1.5 Gyr depletion time scale for massive disks on the blue sequence of star forming galaxies (henceforth ‘main sequence’ (MS): Schiminovich et al. 2007, Noeske et al. 2007, Daddi et al. 2007). From these galaxy integrated data Saintonge et al. (2012) find a global molecular KS- relation of slope N=1.18 (± 0.24). They observe structure within the scatter around this relation, with galaxies having low (high) stellar mass surface densities lying systematically above (below) the mean relation, suggesting that $\Sigma_{H2}$ is not the only parameter controlling the global star formation in galaxies. On scales less than a few hundred parsecs, the KS-relation breaks down in spatially resolved observations in the Milky Way (Murray 2011) and in external galaxies (Onodera et

[2] in all these studies and throughout this paper it is assumed that the conversion factor from CO 1-0 luminosity to molecular gas mass is the same as in the Milky Way ($X_{CO}$= 2x10$^{21}$ cm$^{-2}$/(K km/s), or $\alpha_{CO}$=4.36 $M_\odot$/(K km/s pc$^2$) (e.g. Genzel et al. 2010, Bolatto, Wolfire & Leroy 2013). This value and the resulting molecular depletion time scale include an upward 1.36 correction of the molecular hydrogen masses for the presence of helium.

al. 2010, Schruba et al. 2010, Calzetti, Liu, G & Koda 2012), presumably because of evolutionary effects.

Recently it has become possible to carry out the first systematic studies of the molecular gas content and the molecular KS-relation in massive z>1 star forming galaxies, near or slightly above the MS (henceforth called SFGs or MS SFGs, Tacconi et al. 2010, Daddi et al. 2010a,b, Genzel et al. 2010). In the PHIBSS survey Tacconi et al. (2013) find N=1.05±0.17 from galaxy integrated measurements in 50 z=1-1.5 MS SFGs, and deduce an average depletion time scale of 0.7±0.1 Gyr.

The next step is to explore whether a near-linear high-z KS-relation also applies on sub-galactic scales. In the present paper, we report an analysis of the distribution of molecular gas, extinction, stars, and star formation rate in a very massive z~1.5 MS SFG within the PHIBSS sample, for which we have assembled, for the first time, all the information to carry out a spatially resolved analysis of the z>1 KS-relation.

Throughout the paper, we use the standard WMAP ΛCDM cosmology (Komatsu et al. 2011) and a Chabrier (2003) initial stellar mass function (IMF).

# 2. Observations

## *2.1 The "Plateau de Bure high-z blue-sequence survey" (PHIBSS)*

The PHIBSS survey (Tacconi et al. 2010, 2013) explores the cold molecular gas from $^{12}$CO 3-2 line emission in massive ($\log M_* > 10.4$) MS SFGs at z=1-3. The targets have been selected from parent samples found in large UV/optical/IR look-back imaging surveys, mainly in the AEGIS (DEEP2/DEEP3: Davis et al. 2007, Cooper et al. 2012), GOODS-N (Daddi et al. 2010a, Magnelli et al. 2012) and Caltech QSO fields (Steidel et al. 2004, Erb et al. 2006). With stellar mass and star formation rate cuts of $\log M_* > 10.4$ and SFR≥30 $M_\odot yr^{-1}$ PHIBSS provides a fair census of the massive tail of the star forming population in the stellar mass-star formation rate plane along the MS line and upward, at the peak epoch of cosmic star formation. The MS population accounts for ≥90% of the cosmic star formation rate at z~1-2.5, and as such samples 'normal' massive star forming galaxies near equilibrium growth, rather than rare major merger induced starbursts (Noeske et al. 2007, Daddi et al. 2007, Reddy et al. 2005, Rodighiero et al. 2011, Elbaz et al. 2011, Nordon et al. 2012).

### *2.1.1 EGS13011166*

EGS13011166 (z=1.53, $\log M_*=11.1$) is located in the massive tail of the z~1-2 MS SFGs, and about 0.2 dex above its center line (left panel of Figure 1). On the ACS and WFC3 V-H images of the AEGIS survey field the galaxy is characterized by a very clumpy and amorphous rest-frame UV- and optical light distribution, with a dozen kpc-scale bright clumps distributed over ~5" (~40 kpc, right panel of Figure 1). Table 1 gives a summary of the salient parameters of the galaxy.

We selected EGS13011166 for high resolution follow-up imaging spectroscopy in CO 3-2 and Hα, for the following reasons:

- the large star formation rate and large extent made sub-arcsecond resolution CO observations with the A-configuration of the PdBI feasible,
- the source is at a redshift (z=1.53) where Hα, redshifted into the H-band, is relatively unaffected by strong OH night-sky emission lines,

- Hubble Space Telescope (HST) ACS and CANDELS imagery (Koekemoer et al. 2011, Grogin et al. 2011) in the V, I, J and H bands are publicly available,
- the clumpy UV/optical appearance of EGS13011166 is quite typical for many other z~1.5-2.5 SFGs, and we wanted to test whether the gas kinematics is more regular, and whether these clumps are also visible in molecular emission.

## *2.2 IRAM PdBI CO observations and data analysis*

EGS13011166 was observed for 31 hours in the winter 2012 in the A, B and C configurations of the 6x15m IRAM Plateau de Bure Millimeter Interferometer (Guilloteau et al. 1992, Cox et al. 2011). We observed the $^{12}CO$ 3-2 rotational transition (rest frequency 345.998 GHz), which is shifted into the 2mm band. The observations take advantage of the new generation, dual polarization receivers that deliver receiver temperatures of ~50 K single side band (Cox et al. 2011). Depending on the tapering and configurations used the synthesized beam had a FWHM between 0.7” × 0.85” (for natural weighting in the most sensitive combination A+B+C, rms per channel 0.25 mJy) and 0.56” × 0.68” (in A+B).

System temperatures ranged between 100 and 200 K. Every 20 minutes we alternated source observations with a bright quasar calibrator within 15 degrees of the source. The absolute flux scale was calibrated on MWC349 ($S_{3mm}$=1.2 Jy). The spectral correlator was configured to cover 1 GHz per polarization. The data were calibrated using the CLIC package of the IRAM GILDAS software system and further analyzed and mapped in the MAPPING environment of GILDAS. Final maps were cleaned with the CLARK version of CLEAN implemented in GILDAS. The absolute flux scale is better than ±20%. The spectra were analyzed with the CLASS package within GILDAS.

To convert the measured CO 3-2 fluxes to molecular gas masses (including a 36% upward correction for the presence of helium) we used a Milky Way conversion factor ($\alpha$=4.36 $M_{\odot}$/(K km/s $pc^2$), see footnote 2) and adopted a Rayleigh Jeans brightness temperature ratio of the 1-0 to the 3-2 line of 2 (see Tacconi et al. 2013 for details).

The astrometry of the PdBI maps is accurate to ±0.2", which is consistent with the good spatial alignment of the CO map and the extinction map inferred from the HST imagery and discussed in section 3.1 below.

### *2.3 LUCI@LBT Hα observations and data analysis*

We obtained optical spectroscopy data of EGS13011166 with LUCI (Ageorges et al. 2010, Seifert et al. 2010, Buschkamp et al. 2010) mounted at the Gregorian focus of the 2×8.4m Large Binocular Telescope (LBT) on Mt.Graham in Arizona (Hill et al. 2006). LUCI is a NIR camera and spectrograph that allows multi-object spectroscopy (MOS) at spectral resolving powers of R~3900 over a 2.5'×4' field using user-defined masks.

The data were taken over six consecutive nights in March 2012. We obtained spectra were through a 4'× 0.75" wide slit oriented north-south (cut in a mask), with 0.25" pixels along the slit. We scanned the galaxy in the east-west direction to obtain spatially resolved spectra both along and perpendicular to the slit. We used nine steps of half the slit width (i.e. 0.375") to obtain Nyquist sampling over an extent of 3.4".

The acquisition was done by centering the slit on a star 80” away from the galaxy. In contrast to integral field spectroscopy, where the spectra at all spatial elements are obtained simultaneously, the slit scanning technique obtains spectra at different positions asynchronously and therefore possibly under different atmospheric circumstances. For that reason, we did not stare at one position before moving to the next, but cycled, typically, through five different slit positions in one night. In addition, an offset of 10” along the slit was applied between each 5 minute exposure, following an ABAB pattern. As the seeing was rather stable during the observing run, the PSFs obtained in each slit are similar. After each night, we evaluated the signal-to-noise obtained and decided which positions needed more exposure. This led to inhomogeneous exposure times over the field of view, but improved the signal-to-noise obtained in the outskirts of the galaxy. We, nevertheless, always obtained exposures of the (central) position centered on the acquisition star to monitor its PSF and position on the detector. The number of exposures per slit position varied from 5 to 33, amounting to a total exposure time of 15 hours.

The observations were reduced with the pipeline developed at MPE for LUCI (Kurk et al. in preparation). The pipeline includes removal of bad pixels and persistence effects, cosmic ray masking, wavelength calibration employing OH sky lines, rectification and linearization, background subtraction and co-addition of the exposures. We flux calibrated on an A0V star. We did not attempt to correct for (differential) slit losses. The nine reduced slit positions were co-registered to obtain data and associated noise cubes of the galaxy and the star. The latter has a FWHM of 0.7” along and 0.8” across the slit direction. The spectral resolving power achieved near the Hα line at $\lambda$=1.6610 μm is R~3900 FWHM, as measured from nearby OH lines.

The a priori absolute astrometry of the LUCI data is probably no better than ±0.7”. However, by extracting a continuum H-band map from the H-band cube, we were able to co-align the Hα map on the HST H-band image, to an accuracy of ±0.35”.

### *2.4 HST maps of stellar mass surface density, star formation rate and extinction*

We derived resolved high resolution HST maps of the effective V-band screen extinction, and of the extinction corrected stellar mass and star formation rate surface density distributions within EGS13011166, following the procedures outlined by Förster Schreiber et al. (2011) and Wuyts et al. (2011b, 2012). Briefly, the available multi-wavelength HST imaging was matched to the resolution of the F160W image (0.18” FWHM), employing the PSFMATCH algorithm in IRAF. A Voronoi 2D binning scheme (Cappellari & Coppin 2003) was applied in order to achieve a minimum signal-to-noise level of 10 per spatial bin in the F160W band. Subsequently, we fitted Bruzual & Charlot (2003) stellar population synthesis models to the photometry of each spatial bin, adopting assumptions that are standard in the literature. We allowed ages since the onset of star formation between 50 Myr and the age of the universe, and visual extinctions in the range $0 < A_V < 4$ with the reddening following a Calzetti et al. (2000) recipe. We assume the metallicity of the stellar population is solar across the galaxy (see 3.1.3 below), and allow exponentially declining star formation histories with e-folding times down to $\tau$ =300 Myr. In addition, we also explored delayed tau models (SFR ~ t * exp(-t/$\tau$)), which at early times ($t < \tau$) feature rising star formation rate levels, and find similar results. For comparison to the other data discussed in this paper, the HST images were smoothed with a Gaussian kernel of the desired final FWHM.

# 3. Results

The data presented in this paper allow, for the first time, a spatially resolved comparison of the distribution and kinematics of the molecular and ionized gas (FWHM resolution ~6 kpc), as well as the distribution of the star formation rate, extinction and stellar surface density (FWHM resolution ~1.6-2.5 kpc), in a luminous active MS SFG at the peak of the galaxy formation epoch. We begin with a qualitative comparison of the individual tracers, followed by a quantitative comparison of the kinematics, the extinction/molecular gas maps and the star formation rate/molecular gas maps (see Table 1 for a summary of the basic inferred parameters).

## *3.1 Comparison of the different tracers*

### *3.1.1 Rest-frame UV/optical continuum maps*

The clumpy and irregular rest-frame UV/optical light distribution of EGS13011166 (right panel in Figure 1) is similar to many other z>1 SFGs (Cowie et al. 1996; Giavalisco et al. 1996, van den Bergh et al. 1996; Elmegreen et al. 2005, 2007; Elmegreen & Elmegreen 2006; Förster Schreiber et al. 2009, 2011, Law et al. 2012, Wuyts et al. 2012). The main body of the galaxy is fitted by an $n_S$=1±0.5 Sersic profile in the I/J/H light profiles, with an effective radius $R_{1/2}$=6.3±1 kpc and inclination 60±15$^0$. In addition there are 5 additional prominent clumps outside that main body and distributed over ~40 kpc (labeled A through E in the right panel of Figure 1).

There is an east-west color gradient, with the eastern half of the central galaxy much bluer than the western half, perpendicular to the major axis of the galaxy (p.a. -36$^0$ west of north). This color gradient is largely due to extinction being much higher in a ridge on the western side, as demonstrated in the extinction map derived from the SED fitting to the HST imaging photometry (bottom right panel in Figure 2). The western half of the main galaxy might be in front, such that there is a large column of dust towards the central regions along that line of sight. The centroids of the stellar density and star formation rate are embedded in and slightly east of this dust extinction lane (center and right panels in the top row, as well bottom right panel of Figure 2).

The stellar surface density distribution inferred from the SED fitting is much smoother than the V-H light distributions, consistent with the clumpy distribution of dust, and similar to many other z~1-2.5 SFGs (Wuyts et al. 2012). EGS13011166 thus fits in with the general result of Wuyts et al. (2012) that the rest-frame UV clumps are mostly sites of low extinction star formation.

### *3.1.2 CO*

The molecular surface density map inferred from the CO flux map correlates quite well with the inferred $A_V$-map (bottom central panel of Figure 2), and also with the stellar density (top central panel) and extinction corrected star formation rate (top right panel) maps. However, the brightest CO peak is ~0.5" southeast of the kinematic center (see below) and the stellar and star formation maxima inferred from the extinction corrected UV/optical data. It is unclear whether all or part of this shift is due to a real asymmetry of the molecular gas relative to the extinction distribution, whether it is due to incorrect astrometric alignment, or whether the intrinsic UV/optical emission toward the CO maximum is totally blocked by very large extinction. As we will discuss below, the total dust columns connected to the CO peak correspond to $A_V$~50; unless this dust is distributed in a clumpy arrangement, any underlying UV/optical emission could in principle be completely blocked. Again, the brightest V-/I-band clumps appear to be mostly east (and one west) of this clumpy ridge of molecular column density.

The strong impact of clumpy extinction and dense clouds on the sub-galactic UV/optical light distribution is not surprising given the large gas and dust columns. Similar effects are also seen in z~0 starbursts (e.g. Sams et al. 1994, Satyapal et al. 1997).

### *3.1.3 Optical emission lines*

Figure 3 summarizes the properties of the strong optical emission lines in EGS13011166, as extracted from the LUCI data set. The integrated Hα map (upper right) correlates well with the I/V-band stellar light distributions, but peaks near the prominent CO maximum, south-east of the H-band, stellar surface density maxima and the kinematic center. In contrast the $[NII]_{6585}$ brightness distribution is more compact and peaks very close to the stellar peak (middle right), such that the [NII]/Hα ratio drops somewhat from 0.45±0.05 near the center to 0.3±0.1 ~1" off-center, along the major axis of the disk. The galaxy integrated [NII]/Hα flux ratio of 0.37±0.01 corresponds to a near solar oxygen abundance, 12 +log{O/H}= 8.65 on the Pettini & Pagel (2004) scale, and 8.77 on the Denicolo, Terlevich & Terlevich (2002) scale.

The combined galaxy integrated values for [NII]/Hα and [SII]/Hα can be well accounted for by photo-ionization of slightly super-solar metallicity gas with a modest ionization parameter of logU~-3.2±0.2 (bottom left panel, Newman et al. 2012a). This value is broadly consistent with the 'homogeneous' ionization parameter $\log U < \log(Q_{Lyc} /(4\pi R_{1/2}^2 n_e c)) \sim -3.5 - \log n_{100}$ for the flux of Lyman continuum photons of $Q_{Lyc} \sim 8x10^{43}$ $s^{-1}$ from the extinction corrected Hα luminosity, $L_{Lyc}$=16 $L_{H\alpha}$ and $R_{1/2}$=6.3 kpc, and for an electron density in units of 100 $cm^{-3}$.

The [NII]/Hα gradient could then be caused by a shallow abundance gradient, with 12+log{O/H}~8.55 in the outer disk and 8.73 in the nucleus (-0.02 dex/kpc). Given the relatively low resolution of the LUCI data, this value is likely an upper limit and the intrinsic gradient could be steeper. Alternatively, the maximum of [NII]/Hα at the stellar surface density peak could be due to an AGN there. While the nuclear [NII]/Hα ratio can be well explained by photo-ionization of solar metallicity gas (see above), we cannot exclude that this is strongly affected by beam-smearing and that at high resolution the nuclear [NII]/Hα value may exceed the stellar 'photoionization limit' of ~0.55. There is, however, no X-ray point source associated with EGS13011166 in the Chandra X-ray catalog of AEGIS (http://astro.ic.ac.uk/content/aegis-x, PI K.Nandra).

### *3.1.4 Evidence for a galactic outflow*

The upper left panel of Figure 3 shows the galaxy integrated Hα, [NII] and [SII] spectra. The line profile between Hα and the two [NII] lines does not dip to the zero level, and there are non-Gaussian wings blue- and red-ward of the two [NII] lines. The spectral profiles require a strongly non-Gaussian shape or, more likely, a second, underlying broad emission component (see discussions in Shapiro et al. 2009, Genzel et al. 2011, Newman et al. 2012b). A global two-Gaussian fit to all five emission lines, indicates that the broad emission is present in Hα and the [NII] lines, and perhaps also in the [SII] lines.

These characteristics are seen commonly in z>1 MS SFGs (Shapiro et al. 2009, Genzel et al. 2011, Newman et al. 2012b). Since the broad emission is observed in the forbidden lines as well, it is probably caused by an extended ionized gas outflow, and not by a compact, AGN broad line region. The width of the broad emission, $\Delta v_{FWHM}$~700 km/s, suggests an average outflow velocity of $v_{out}$~600 km/s (Genzel et al. 2011), which is comparable to spatially resolved ionized outflows most likely driven by star formation activity ('star formation feedback') seen in many other z~1-2.5 SFGs (Newman et al. 2012b). AGN driven winds result in yet broader emission ($\Delta v_{FWHM}$~1400 km/s); such winds are detected toward the nuclei of a number of logM*>11 SFGs (Förster Schreiber et al. 2013 in prep.). The spatial resolution of the seeing limited LUCI data is not sufficient to distinguish between these two possibilities but the line width is more suggestive of 'star formation' feedback.

From the measured line width and broad to narrow flux ratio $F_{broad}/F_{narrow}$= 0.48±0.13, a mass outflow rate can be estimated ($\dot{M}_{out} \propto L(H\alpha)_{narrow} \times \frac{F_{broad}}{F_{narrow}} \times \frac{v_{out}}{n_e R}$, Genzel et al. 2011). We assume a source radius $R$ comparable to the half-light radius of the galactic disk, and an electron density of $n_e$~50 cm$^{-3}$ (Newman et al. 2012b), motivated by the fact the broad $[SII]_{6718}/[SII]_{6732}$ ratio is close to the low density limit. We infer a mass outflow rate of $\dot{M}_o = 570\ M_\odot yr^{-1}$, and a mass loading factor $\eta = \dot{M}_o / SFR$ = 1.5±1. While obviously quite uncertain, the inferred mass loading parameter is in good agreement with the results of Newman et al. (2012b) who find that z~1.5-2.5 SFGs exhibit powerful winds with $\eta$~2 above a star formation rate surface density threshold of ~1.5 $M_\odot yr^{-1} kpc^{-2}$. The star formation rate surface density of EGS13011166 is $<\Sigma_{star\ form}>=0.5\times SFR/(\pi R_{1/2}^2)$~1.5 $M_\odot yr^{-1} kpc^{-2}$.

### *3.2 Gas kinematics and modeling*

The main body of the galaxy exhibits a regular velocity field in CO (Figure 4) and Hα (Figure 6), with a progression from blue-shifted emission (-215 km/s) in the south-east to red-shifted emission (+210 km/s, both relative to z=1.5307), along the main body of the galaxy. At our nominal registration, the kinematic centroid and the peak of beam-smeared velocity dispersion are ~0.4" ESE of the extinction corrected stellar and star formation rate peaks. This offset is marginally significant, given the combined relative astrometric uncertainties. The iso-velocity contours in the middle left map of Figure 4 exhibit the characteristic 'spider' shape of a rotating disk. The overall velocity and velocity dispersion fields in CO (and Hα, Figure 6) are well fitted by a rotating disk with an exponential half-light radius comparable to that of the UV/optical light distribution ($R_{1/2}$(CO)=6.3±1 kpc). The inclination corrected, maximum rotation velocity is 307±50 km/s.

After removal of the rotation and instrumental broadening the average residual dispersion in the outer parts of the galaxy (where the contribution from the residual beam-smeared rotation is smallest) is $\sigma_0$=52±10 km/s. The resulting ratio $v_{rot}/\sigma_0$ is 6.1±2. This is illustrated in the central and right columns of Figure 4, where we show an exponential model disk, and the residuals between data and model. Considering the residual velocity map in the middle right panel of Figure 4, the average residual is only 18 km/s. It needs to be kept in mind, however, that the model has seven free parameters (centroid coordinates, systemic velocity, total dynamical mass, inclination, position angle of the line of nodes, radial scale length of exponential distribution) that we varied to obtain the results in Figure 4. Our good fit result, therefore, is not a unique solution.

Including a correction for the pressure term (Burkert et al. 2010) we infer a total dynamical mass of 3±0.6 $\times 10^{11}$ $M_{\odot}$ within ~2.5" (21 kpc). For comparison, the total stellar mass from galaxy integrated SED fitting is 1.2±0.4$\times 10^{11}$ $M_{\odot}$, and the total gas mass, as inferred from the integrated CO flux is 2.6±0.8×($\alpha_{CO}$/4.36)$\times 10^{11}$ $M_{\odot}$ (Tacconi et al. 2013). The ratio of baryonic to dynamical mass thus is 1.26±0.6, fully consistent with a baryon dominated system. Given the large systematic uncertainties in all mass tracers, there could also be a substantial dark matter contribution.

### *3.2.1 Toomre Q-parameter*

With a model of the intrinsic rotation curve and an estimate of the intrinsic velocity dispersion (assumed for simplicity to be spatially constant and isotropic), we combine the kinematic parameters and the molecular surface density distribution and compute the Toomre parameter (Toomre 1964) of the EGS1301166 disk,

$Q_{Toomre} = \frac{\kappa \sigma_0}{\pi G \Sigma_{molgas}}$. Here κ is the epicyclic frequency,

$\kappa^2 = R \times d((v/R)^2)/dR \ + 4 \times (v/R)^2$, $v$ is the rotation velocity at radius $R$, $\sigma_0$ the average local velocity dispersion and $\Sigma_{molgas}$ is the molecular gas surface density. Figure 5 shows the inferred Q-distribution along the major axis of the galaxy, in a 0.75" software slit, superposed on the molecular surface density distribution. With the assumptions made above, the Toomre Q parameter is below unity throughout the disk, although the exact absolute value is uncertain by at least ±0.3dex, given the combined systematic uncertainties entering the computation of Q. The disk is thus globally unstable to gravitational fragmentation, at least averaged over our ~6 kpc resolution, and consistent with findings in several other z~1-2 systems (Genzel et al. 2006, 2011).

*3.2.2 Central velocity dispersion & evidence for star forming bulge*

We observe significant kinematic anomalies in the data-minus-model residual maps of Figure 4. The largest anomalies are in the center of the galaxy in the dispersion residual map, and in the southern extension of the galaxy in both velocity and velocity dispersion residual maps (in both CO and Hα data, see below 3.2.3). The central maximum in the residual velocity dispersion is significantly larger than expected from the beam-smeared rotation in an exponential disk galaxy with inclination $60^0$. Reducing the inclination decreases the beam-smeared velocity dispersion residual somewhat (because of the assumption of isotropic velocity dispersion in a thick disk) but does not eliminate the effect, unless an unrealistically low inclination is adopted.

A more likely explanation is a mass profile that is more concentrated mass distribution than the exponential disk profile assumed in the modeling of Figure 4. Such a mass concentration would be consistent with the presence of a substantial bulge component, which is also suggested qualitatively by the appearance of the de-reddened stellar surface density map (bottom right panel in Figure 2). An $R_{1/2}$=6.3 kpc exponential disk of total mass $3x10^{11}$ $M_{\odot}$ (gas and stars) would have $2x10^{10}$ $M_{\odot}$, or 7% of its mass, within the central R≤1.5 kpc (0.18”). The inferred stellar mass fraction within that radius deduced from the bottom right panel of Figure 2 is twice as large, or 14%. Within 0.4” (3.4 kpc) the exponential model would have 23% of the total mass, while the inferred stellar mass is 36%, which places EGS13011166 among the ‘mature bulgy’ systems in the $M(\leq0.4'')/M_{tot}$ versus [NII]/Hα diagram of Genzel et al. (2008). Assuming that the mix of dust and gas is radially constant, this corresponds to a total baryonic mass of ~$4 \times 10^{10}$ $M_\odot$ within R=1.5 kpc. For an isothermal distribution this mass would result in a velocity dispersion of 240 km/s at that radius, in agreement with the observed dispersion in the bottom left panel of Figure 4. We conclude that EGS13011166 has a substantial nuclear bulge. Since the dereddened star formation rate peaks there as well, the bulge is actively star forming.

### *3.2.3 Evidence for minor merger & clumps*

There is a significant extension of the main body of the galaxy to the south, at a radius of ~19 kpc, visible in the UV/optical, Hα and CO maps, and with two prominent embedded clumps (labeled 'C' and 'D' in Figure 1). This velocity of this component is offset by +130 km/s with respect to the extrapolated disk velocity in this direction. It is thus quite likely that the southern component is due to an ongoing interaction or minor merger. If so, the companion has a 1:10 mass star formation rate ratio with respect to the main galaxy, as derived from the spatially resolved SED modeling. The ratio of molecular gas mass to star formation rate of this southern companion is similar to that of the main galaxy, corresponding to a depletion time scale of 0.3-1 Gyr, consistent with a normal star forming galaxy.

There are three other prominent UV/optical clumps north, east and south-east of the galaxy, at similar radii (Figure 1). CO is only marginally detected in the eastern clump ('B') at a velocity of -410 km/s relative to systemic, while CO emission spikes towards 'A' and 'E' are not statistically significant. It is thus not clear whether these clumps are part of the EGS13011166 system.

### *3.2.2 Comparison of CO and Hα kinematics*

The line profiles, velocity channel maps, major axis velocity and velocity dispersion profiles of the CO and Hα data are compared in Figure 6. They exhibit a remarkable agreement. This is notwithstanding the facts that the Hα integrated map and the channel maps exhibit a tendency for the near-systemic Hα emission to be located east of the CO emission at these velocities, and for the south-eastern CO maximum to be less prominent and further north-west in the Hα data. These finding are plausibly consistent, however, with the overall east-west color and extinction gradients discussed in 3.1.

The agreement in molecular and ionized gas kinematics supports the interpretation that, at least in this massive SFG, both tracers can be equally well used for establishing the rotation curve and the galaxy kinematics. Where measurable with significance, the very good agreement of the intrinsic CO and Hα velocity dispersions in the outer parts of the galaxy along the major axis, where beam smearing effects should be minimum, supports the general conclusion mainly from Hα-work that the interstellar gas layer in high-z SFGs is highly turbulent ($\sigma_0$~50-60 km/s), and high-z disks thus are necessarily geometrically thick ($h_z/R \sim \sigma_0/v_{rot} \sim 0.15$-0.3, Genzel et al. 2006, 2011, Förster Schreiber et al. 2009). The results presented here for a resolved case suggest that this conclusion holds for the entire star forming ISM, and not just its ionized component (see also Tacconi et al. 2013 for several other cases in PHIBSS).

### 3.3 The spatially resolved molecular gas-star formation rate relation

As explained in the Introduction, our primary motivation for collecting such detailed CO, Hα and HST observations in this galaxy was to study quantitatively the z>1 molecular gas to star formation rate relation with spatially resolved data. The galaxy integrated gas depletion time scale of EGS13011166, $\tau_{depl}$= $M_{mol\ gas}$/SFR=0.69 Gyrs (Table 1), is identical with the average depletion time scale of z~1-1.5 MS SFGs (Tacconi et al. 2013).

To explore the spatially resolved KS-relation, we grid the CO and HST data on a 0.25”×0.25” pixel scale, oversampling the original LUCI Hα data by 50%. In Figure 7 we compare the inferred molecular gas surface densities (from Gaussian fitting to the CO data) to the effective screen extinction values inferred from the HST spatially resolved SED modeling, and smoothed to the same 0.75” FHWM resolution. The left panel of Figure 7 shows the overlay of these maps, with good qualitative agreement between the inferred optical extinction and the total molecular column, as noted before in section 3.1.

The blue circles in the right panel of Figure 7 denote how these values compare quantitatively pixel by pixel. We note that the correlation of the data points is significantly better than what would be expected from the $A_V$ error bars obtained from pixel by pixel broad-band continuum SED fitting (large cross). This is partly due to the correlation of sub-Nyquist-sampled data. More importantly, the extinction uncertinaties include the impact of systematics stemming from varying star formation histories in the fitting procedure etc. and thus – in a relative sense – are probably too conservative.

The observed correlation is non-linear. It can be quite well modeled in terms of the conversion of a fully mixed distribution of gas and dust into the framework of an effective screen, where one would expect (e.g. Wuyts et al. 2011a)

$$\frac{I(H\alpha)_{obs}}{I(H\alpha)_{intr}} = \frac{1-\exp(-\tau_{tot}(H\alpha))}{\tau_{tot}(H\alpha)} \equiv \exp(-\tau_{eff}(H\alpha)) \qquad (3)\ ,$$

where $\tau_{eff}$ and $\tau_{tot}$ are the effective screen and total dust optical depths along a given sightline. The total optical depth is related to the molecular column density through

$$\tau_{tot}(H\alpha) = \frac{\Sigma_{molgas}(M_{\odot}\,pc^{-2})}{g \times \Sigma_{MW}} \qquad (4),$$

where $\Sigma_{MW}$=30 $M_{\odot}pc^{-2}$ is the hydrogen column density of gas that is equivalent to a dust optical depth of unity in the R-band in the diffuse ISM of the Milky Way (at solar metallicity, N(H)/$A_V$~2x$10^{21}$ $cm^{-2}$/mag,  Bohlin et al. 1978). The dimensionless factor g then denotes how much this relation deviates empirically from the MW-relation. The best fit value of equation (3) for the data in Figure 7 yields a value of g~4.8±0.5. That is, the data in EGS13011166, in terms of the mixed model in equation (3), require a total dust to gas column ratio ~5 times less than in the diffuse ISM of the Milky Way. The fact that the data in Figure 7 seem to flatten even more than equation (3) can accommodate, might suggest that g increases with $\Sigma_{molgas}$.

Nordon et al. (2013) have recently investigated the relation between the ratio of far-infrared to UV-flux densities and the UV SED slopes $\beta_{UV}$ in high-z SFGs. They also investigated the correspondence of these values to the total molecular columns as inferred from CO observations. Nordon et al. find from this totally independent approach that massive high-z SFGs are well described by a mixed dust/gas model with a ratio of effective UV-extinction to total UV column about 5 times smaller than expected from diffuse Milky-Way ISM, which can be seen from the location of solar-metallicity main-sequence SFGs in the $A_{IRX}$-$\beta_{UV}$ plane (lower panel of their Fig.10, yielding $\Delta A/A_{IRX}$~5). This result is in good agreement with our findings.

### *3.3.1. KS-slope for a 'mixed' extinction correction*

The success of this comparison encourages us to correct the observed Hα distribution, converted to a star formation rate, pixel by pixel with the $A_V$-map derived from the broad-band SED fitting (using $\tau_{H\alpha\ eff}$=0.73×$A_{V\ eff}$, Calzetti et al. 2000). Likewise one can also use the molecular surface density distribution and then use equations (3) and (4), with similar results. The resulting pixel-by-pixel relation between molecular gas and star formation rate surface densities is shown as green open squares (with 1σ uncertainties) in the left panel of Figure 8. Black filled circles give weighted averages (and their dispersions) of the same data, binning the green points in groups of 10-30. In following up on the likely overestimate of the $A_V$-errors in the SED fitting, we have assumed here, that the (relative) $A_V$ errors are half those estimated from the SED fitting. A weighted fit of the 110-170 data points in this plot, depending on choices of significance cutoffs etc., yields a slope of N=1.14. The statistical fit uncertainty is ±0.05 and the systematic uncertainty is about ±0.1…±0.15. With this modeling, the slope of the z~1.5 molecular KS-relation is slightly above unity, in good agreement with the galaxy integrated analysis presented in Tacconi et al. (2013) for 50 z=1-1.5 SFGs, and with the radial aperture analysis of Freundlich et al. (2013) in 4 z=1-1.5 SFGs.

### *3.3.2 KS-slope for different extinction correction methods*

The inferred KS-slope depends strongly on the method of the extinction correction applied to the spatially resolved data. This is demonstrated in the right panel of Figure 8, where the binned weighted averages are shown for different extinction correction methods,

1) the 'single' Calzetti effective screen extinction correction of the last section (identical with a 'mixed' extinction correction from the total gas column), for each pixel in the observed Hα luminosity map, $SFR_{H\alpha\,0}$ ($M_\odot yr^{-1}$) $=\exp(0.73\times A_{Veff}) \times L(H\alpha)_{obs}/2.1e42$ (erg/s) (K98),
2) a single value ('global') extinction correction (11.4) of the observed Hα luminosity map, converted to star formation rate with $SFR_{H\alpha\,global}$ ($M_\odot yr^{-1}$) $=11.4\times L(H\alpha)_{obs}/2.1e42$ (erg/s) (K98). This method corrects the integrated star formation rate inferred from the observed Hα luminosity (33 $M_\odot$ $yr^{-1}$) to the 'true' total star formation rate, inferred from the sum of the observed UV-luminosity (23 $M_\odot yr^{-1}$), and the Herschel-PACS far-infrared luminosity (352 $M_\odot yr^{-1}$: PEP survey of Lutz et al. 2011, $SFR(UV+IR)=1.09\times10^{-10}\times(L(IR)+3.3\times L(2800))$, Wuyts et al. 2011b),
3) a 'double' Calzetti effective screen extinction correction for each pixel in the observed Hα luminosity map, $SFR_{H\alpha\,00}$ ($M_\odot yr^{-1}$) $=\exp(2.27\times0.73\times A_{Veff}) \times L(H\alpha)_{obs}/2.1e42$ (erg/s) (K98). This method is motivated by the finding in local starburst galaxies that the nebular gas has 2.27 times greater extinction than the stars (Calzetti et al. 2000). And finally
4) a single 'broadband' Calzetti screen extinction correction, as in 1), but for the star formation rate inferred from the broadband UV/optical SED modeling (section 2.4).

In the right panel of Figure 8, the binned results for the method 1), ‘mixed’ correction, are again (as in the left panel) shown as filled black circles (N~1.14). Method 4), the‘broadband’ correction, yields a similar slope of N~1.15 (open blue circles). For method 2), the ‘global’ correction, the slope of the resulting relation is flatter (N~0.8, crossed red squares). Method 3), the ‘double Calzetti’ correction of the Hα data, yields the steepest slope (N~1.7, cyan).

Naturally these different extinction correction methods also result in quite different integrated star formation rates. While the ‘single effective screen/mixed’ correction perhaps is the most plausible method, given Figure 7, it delivers an integrated star formation rate of ~114 $M_{\odot}yr^{-1}$, a factor 3 below the ‘ground-zero’ star formation rate. In contrast, the ‘double Calzetti’ correction yields ~560 $M_{\odot}yr^{-1}$, or a factor 1.5 above the ‘ground-zero’ value. The spatially resolved SED-fitting method yields an even integrated star formation rate of 950 $M_{\odot}yr^{-1}$, a factor 2.5 too large. By design, the ‘global’ correction matches the true value of 375 $M_{\odot}$ $yr^{-1}$.

Clearly spatially resolved data as presented in this paper require a better empirical description of the spatially variable extinction component, which comes from a mixture of dust distributed diffusely throughout the galaxy, as well as dust located in the individual star formation regions. The former will correspond to the ‘single Calzetti/mixed’ correction estimated from the broad-band SED fitting. The latter may perhaps better be described as a constant average extinction (for each star formation region), rather than an extra extinction scaling with the diffuse screen correction, as in the ‘double Calzetti’ recipe (Wuyts et al. in prep., Nordon et al. 2013). In some cases, spatially resolved maps of the Hα/Hβ Balmer decrement may give a direct, spatially resolved measure of the extinction of the nebular gas.

Alternatively one could observe a spatially resolved tracer of the obscured star formation. A ‘true’ star formation estimator then is the sum of this and the observed Hα star formation rate, similar to the methods applied in the local Universe (Kennicutt & Evans 2012, Kennicutt et al.2009, Calzetti et al. 2007, 2010). An obvious tool is submillimeter continuum imaging with ALMA, which corresponds to the rest frame far-infrared dust emission peak in high-z SFGs.

# 4. Summary

We have reported a comprehensive data set on the massive star forming galaxy EGS13011166 near the main star formation sequence at the peak of galaxy formation (z~1.5), combining resolution matched imaging spectroscopy and kinematics of molecular (CO 3-2) and ionized (Hα) gas components, and high resolution 4-band rest-frame UV to optical HST imagery.

- In the short wavelength bands the light distribution is characterized by a highly asymmetric and clumpy appearance spread over almost 40 kpc, similar to many other near-MS z>1 SFGs. UV/optical colors exhibit strong spatial variations and are the result of extinction variations that correlate well with the molecular gas column density distribution inferred from the CO 3-2 map.
- In contrast, extinction corrected stellar mass and star formation rate distributions and the remarkably similar CO/Hα velocity distributions are smoother. The main body of the galaxy appears to be a large ($R_{1/2}$~6.3 kpc), inclined rotating disk, plus a massive (logM~10.6), star forming central bulge.

- The disk has a Toomre Q-parameter below unity and thus is globally unstable to fragmentation on the scales we have been probing.
- The CO/Hα velocity distribution suggests that the southern extension of this disk may be a smaller (1:10) galaxy currently interacting or merging with the massive main system (logM=11.5). A few of the other bright extra-disk clumps may also be part of the same galaxy group.
- The non-linear pixel-by-pixel correlation between the molecular gas column density and the effective extinction, inferred from SED modeling of the high resolution HST data, is quite well modeled by a 'mixed' dust/gas-stars model in which the ratio of total gas to dust column is about 5 times smaller than in the diffuse ISM of the Milky Way.
- The inferred slope of the spatially resolved KS-relation is broadly consistent with those found in the local Universe (N~1-1.5), but the detailed value, and thus also the resulting extinction corrected star formation rate, strongly depends on the extinction model adopted. To overcome this fundamental limitation in future work, it will be necessary to map a direct probe of the nebular extinction (via the Balmer decrement), or to obtain a tracer of the obscured star formation. An obvious tool is submillimeter continuum imaging with ALMA.

**Acknowledgements:** The observations presented here would not have been possible without the diligence and sensitive new generation receivers from the IRAM staff – for this they have our highest admiration and thanks. We also thank the astronomers on duty and telescope operators for delivering consistently high quality data to our team. We also thank the LUCI team and the staff of the Large Binocular Telescope Observatory for their support of these observations. ADB wishes to acknowledge partial support from a CAREER grant NSF-AST0955836, and from a Research Corporation for Science Advancement Cottrell Scholar award. This work is based on observations taken by the CANDELS Multi-Cycle Treasury Program with the NASA/ESA HST, which is operated by the Association of Universities of Research in Astronomy, Inc., under NASA contract NAS5-26555.

# References

Ageorges, N. et al. 2010, SPIE 7735, 53

Bigiel, F., Leroy, A., Walter, F., Brinks, E., de Blok, W. J. G., Madore, B. & Thornley, M. D. 2008, AJ, 136, 2846

Bigiel, F. et al. 2011, ApJ, 730, 13

Bohlin, R.C., Savage, B.D. & Drake, J.F. 1978, ApJ 224, 132

Bolatto, A. D., Wolfire, M. & Leroy, A. K. 2013, ARAA in press

Buschkamp, P. et al. 2010, SPIE 7735, 236

Burkert, A. et al. 2010, ApJ 725, 2324

Bruzual, G. & Charlot, S. 2003, MNRAS 344, 1000

Calzetti, D., Armus, L., Bohlin, R. C., Kinney, A. L., Koornneef, J.& Storchi-Bergmann, T. 2000, ApJ 533, 682

Calzetti et al. 2007, ApJ 666, 870

Calzetti, D. et al. 2010, ApJ 714, 1256

Calzetti, D., Liu, G. & Koda, J. 2012, ApJ 752, 98

Cappellari, M. & Copin, Y. 2003, MNRAS 342, 345

Chabrier, G. 2003, PASP 115, 763

Cowie, L.L., Songaila, A., Hu, E.M. & Cohen, J.G. 1996, AJ 112, 839

Cox, P. et al. 2011, IRAM Annual Reports, http://iram.fr/IRAMFR/ARN/AnnualReports/Years.html

Cooper, M.C. et al. 2012, MNRAS 419, 3018

Daddi, E. et al. 2007, ApJ 670, 156

Daddi, E. et al. 2010a, ApJ 713, 686

Daddi, E. et al. 2010b, ApJ 714, L118

Davis, M. et al. 2007, ApJ 660, L1

Denicoló, G., Terlevich, R. & Terlevich, E.  2002, MNRAS 330, 69

Elbaz, D. et al. 2011, A&A 533, 119

Elmegreen, D. M., Elmegreen, B. G., Rubin, D. S. & Schaffer, M. A. 2005, ApJ 631, 85

Elmegreen, B.G. & Elmegreen, D.M. 2006, ApJ 650, 644

Elmegreen, D. M., Elmegreen, B. G., Ravindranath, S., & Coe, D. A.  2007, ApJ 658, 763

Erb, D. K., Steidel, C. C., Shapley, A. E., Pettini, M, Reddy, N. A. & Adelberger, K. L. 2006, ApJ 647,128

Förster Schreiber, N. M. et al. 2009, ApJ 706, 1364

Förster Schreiber, N. M, Shapley, A. E.,  Erb, D. K., Genzel, R., Steidel, C. C., Bouché, N., Cresci, G. & Davies, R. 2011, ApJ 731, 65

Freundlich, J. et al. 2013, submitted to A&A

Genzel, R. et al. 2006, Nature 442, 786

Genzel, R. et al. 2008, ApJ 687, 59

Genzel, R. et al. 2010, MNRAS 407, 2091

Genzel, R. et al. 2011, ApJ 733, 101

Giavalisco, M., Steidel, C.C. & Macchetto, F.D. 1996, ApJ 470, 189

Grogin, N.A. et al.  2011, ApJS 197, 35

Guilloteau, S. et al. 1992, A&A 262, 624

Hill, J.M., Green, R.F. & Slagle, J.H. 2007, SPIE 6267, 31

Kennicutt R. C., Jr., 1998a, ApJ 498, 541

Kennicutt, R. C., Jr. 1998b, ARA&A 36, 189

Kennicutt R. C., Jr. et al. 2007, ApJ 671, 333

Kennicutt, R.C. et al. 2009, ApJ 703, 1672

Kennicutt, R.C., Jr. & Evans, N. 2012, ARA&A 50, 531

Koekemoer, A.M. et al. 2011, ApJS 197, 36

Komatsu, E. et al. 2011, ApJS 192, 18

Krumholz, M. R. & Tan, J.C. 2007, ApJ 654, 304

Krumholz, M. R., Dekel, A. & McKee, C. F. 2012, ApJ 745, 69

Law, D. R., Steidel, C. C., Shapley, A. E., Nagy, S. R., Reddy, N. A. & Erb, D. K. 2012, ApJ 745, 85

Leroy, A. K., Walter, F., Brinks, E., Bigiel, F., de Blok, W. J. G., Madore, B. & Thornley, M. D. 2008, AJ 136, 2782

Leroy, A.K. et al. 2013, ApJ in press (2013arXiv1301.2328)

Lutz, D. et al. 2011, A&A 532, 90

Magnelli, B. et al. 2012, A&A, 548, 22

Mancini, C., et al. 2011, ApJ 743, 86

McCracken, H. J. et al. 2010, ApJ 708, 202

McKee, C. F. & Ostriker, E. C. 2007, ARA&A 45, 565

Murray, N. 2011, ApJ 729, 133

Newman, S. et al. 2012a, ApJ 752, 111

Newman,S. et al. 2012b, ApJ 761, 43

Noeske K. G. et al. 2007, ApJ 660, L43

Nordon, R. et al. 2012, ApJ 745, 182

Nordon, R. et al. 2013, ApJ 762, 125

Onodera, S. et al. 2010, ApJ 722, L127

Pettini, M. & Pagel, B.E.J. 2004, MNRAS 348, L59

Rahman, N. et al. 2012. ApJ 745, 183

Reddy, N. A., Erb, D. K., Steidel, C. C., Shapley, A. E., Adelberger, K. L., & Pettini, M. A. 2005, ApJ 633, 748

Rodighiero, G. et al. 2011, ApJ 739, 40

Saintonge, A. et al. 2011a, MNRAS 415, 32

Saintonge, A. et al. 2011b, MNRAS 415, 61

Saintonge, A. et al. 2012, ApJ 758, 73

Sams, B. J., III; Genzel, R.; Eckart, A.; Tacconi-Garman, L. & Hofmann, R. 1994, ApJ 430, L33

Satyapal,S. 1997, ApJ 483, 148

Schiminovich, D. et al. 2007, ApJS 173, 315

Schmidt, M. 1959, ApJ 129, 243

Schruba, A., Leroy, A. K., Walter, F., Sandstrom, K. & Rosolowsky, E. 2010, ApJ 722, 1699

Schruba et al. 2011, AJ 142, 37

Seifert,W. et al. 2010, SPIE 7735, 256

Shapiro, K.L. et al. 2009, ApJ 701, 955

Steidel, C. C., Shapley, A. E., Pettini, M., Adelberger, K. L., Erb, D. K., Reddy, N. A. & Hunt, M. P. 2004, ApJ 604, 534

Tacconi, L. J. et al. 2010, Nature, 463, 781

Tacconi, L.J. et al. 2013, ApJ in press (astro-ph arXiv:1211.5743)

Toomre, A. 1964, ApJ 139, 1217

van den Bergh, S. , Abraham, R. G., Ellis, R. S., Tanvir, N. R., Santiago, B. X. & Glazebrook, K. G. 1996, AJ 112, 359

Wuyts, S. et al. 2011a, ApJ 742, 96

Wuyts, S. et al. 2011b, ApJ 738, 106

Wuyts, S. et al. 2012, ApJ 753, 114

# Figures

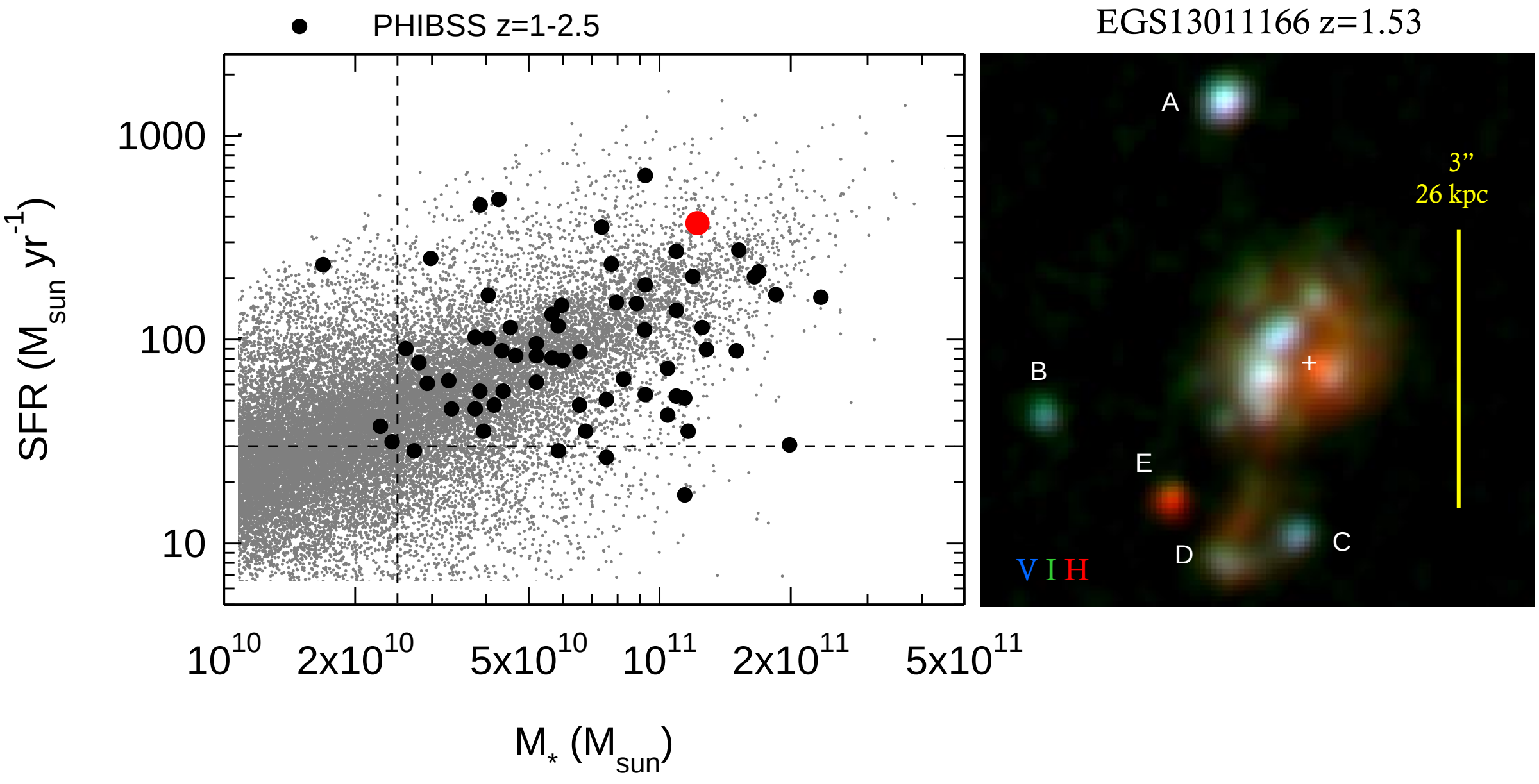

Figure 1. Left: The location of EGS13011166 in the stellar mass – star formation rate plane (filled red circle), compared to the overall PHIBSS CO 3-2 survey of z~1-3 SFGs (filled black circles, Tacconi et al. 2013), and the underlying MS of z~1.5-2.5 SFG population as derived in the COSMOS field with the BzK color-magnitude criterion (grey dots, McCracken et al. 2010, Mancini et al. 2011). Right: V (blue) –I (green) –H (red) – three band HST image of EGS13011166, all on a sqrt(F) color scale. The images were smoothed to ~0.25” FWHM. A-E denote prominent star formation clumps. The small white cross marks the position of the stellar and star formation rate surface density peaks (Figure 2).

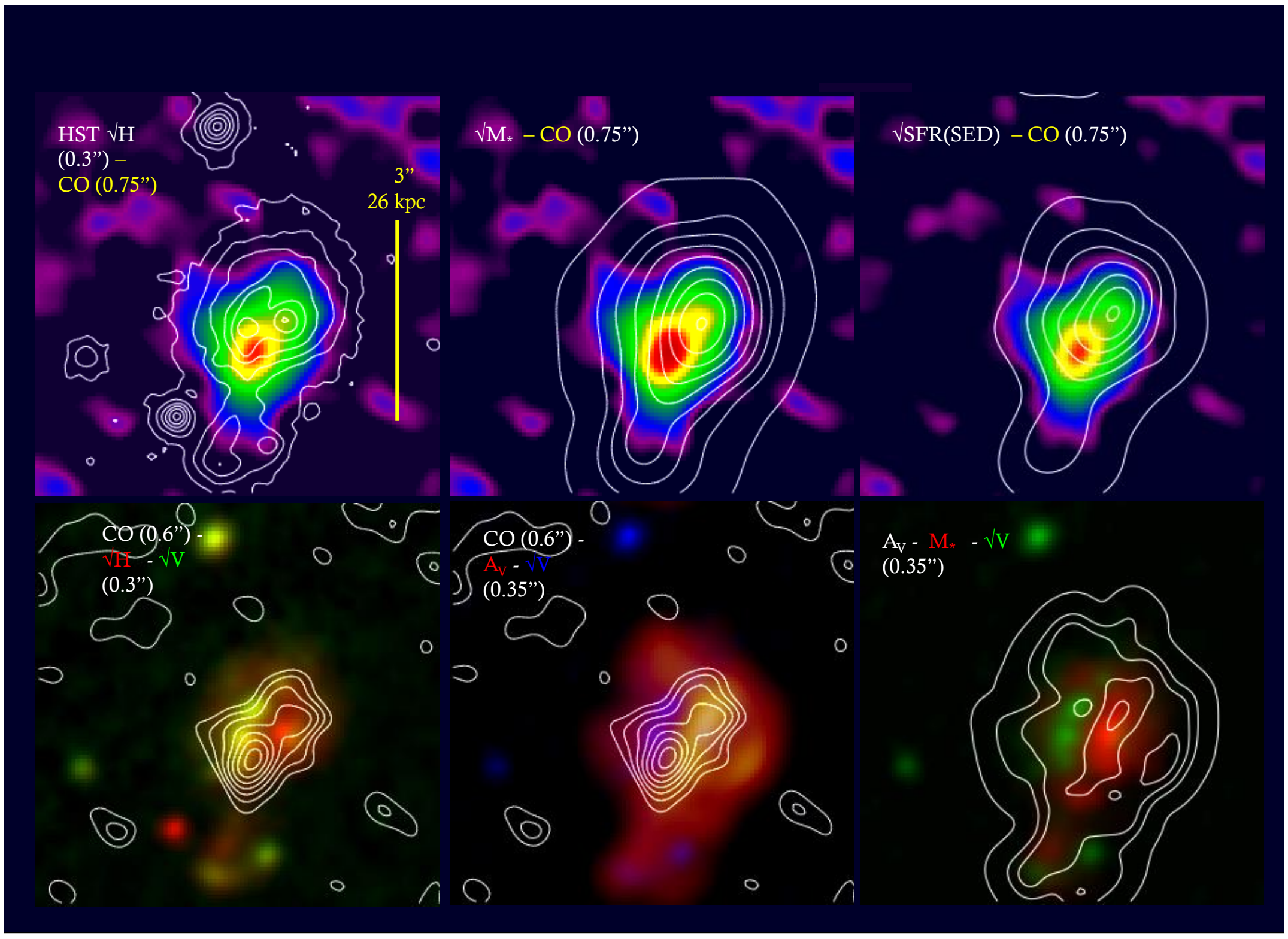

Figure 2: Comparison of images. Top row: integrated CO flux map in color (ABC-configurations, 0.7×0.8" FWHM) and (left) WFC3 HST H-band mapas white contours, ~0.3" FWHM; (center) inferred stellar mass distribution from SED fitting white contours, 0.75" FWHM); and (right) extinction corrected star formation rate distribution inferred from SED fitting (white contours, 0.75" resolution). Bottom row: (left) CO integrated flux map in white contours (AB-configuration, 0.56"×0.68" FWHM) and HST WFC3 H-band map (red, 0.3" FWHM) and ACS V-band map (green, 0.3" FWHM), (center) CO integrated flux map in white contours, (AB-configuration, 0.56"×0.68" FWHM) and HST $A_V$- map (red, 0.35" FWHM) and (right) ACS V-band map (blue, 0.35" FWHM). (right) HST ACS V-band (green), and $A_V$-map in white contours and stellar mass map (red), both inferred from SED modeling. All HST maps are smoothed to 0.35" FWHM. The pixel scale is 0.06".

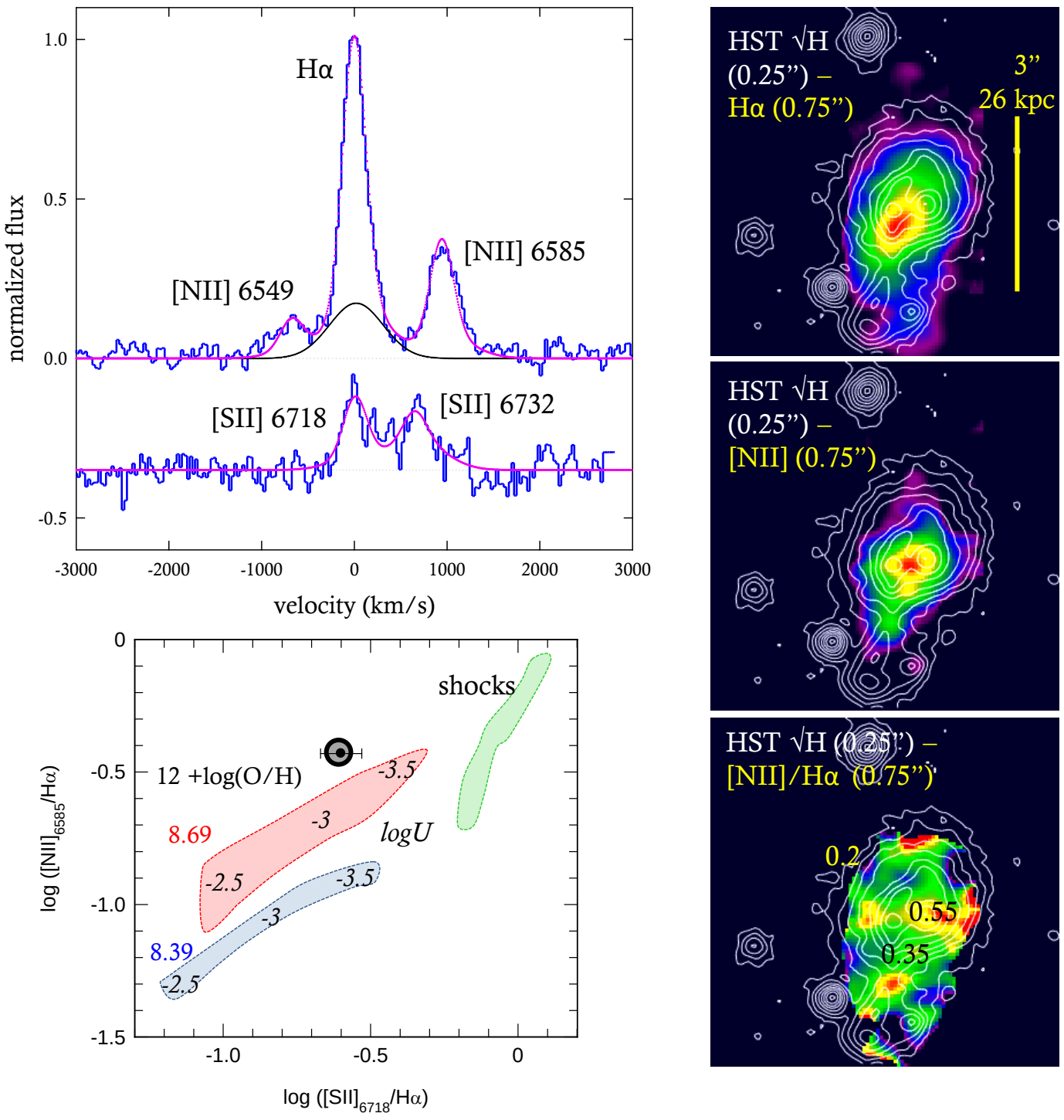

Figure 3. LUCI spectroscopy. Top left: integrated spectrum of EGS13011166, after removing pixel-by-pixel velocity shifts due to the large scale rotation. It shows Hα, [NII] and [SII] emission lines (blue). The pink curve is the best fit double Gaussian profile for all 5 lines: a 'narrow' component of $\Delta v_{FWHM}$=293(±11) km/s, and flux ratios $[NII]_{6585}$ /Hα=0.37±0.01, $[SII]_{6718}$/Hα=0.25±0.04 and $[SII]_{6718}/[SII]_{6732}$~1.3(+0.8,-0.1), and a 'broad' component (black curve) with $\Delta v_{FWHM}$=680 (±80) km/s and flux ratio $(F_{broad}/F_{narrow})$(Hα)=0.48(±0.13). The broad component is also present in the [NII] and [SII] lines. Bottom left: location of ESG13011166 in the [NII]/Hα and [SII]/Hα plane (adapted from Newman et al. 2012a), along with photo-ionization and shock models. The spatially integrated value for EGS13011166 (large black circle) is well fit with solar metallicity photoionization models with ionization parameter logU~ -3.2. Right: Overlays the HST H-band map

(white contours) on the integrated Hα flux map (top), integrated $[NII]_{6585}$ flux map (middle) and $[NII]_{6585}$/Hα flux ratio map (bottom).

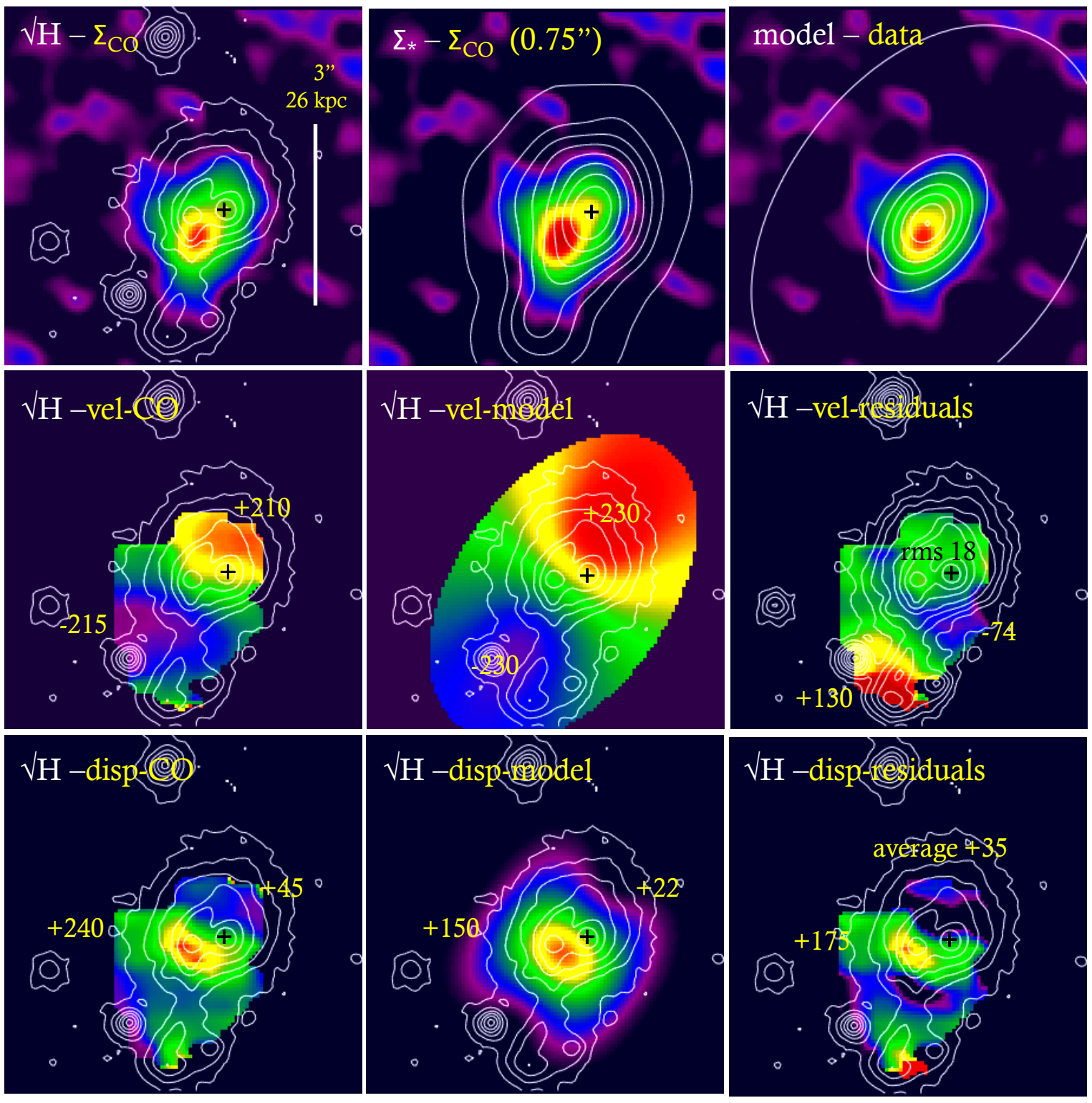

Figure 4: molecular gas derived kinematics and disk models in EGS13011166 (all from ABC-configuration data at 0.7" × 0.8" FWHM resolution). Top row CO integrated flux map in color with white contour maps of (from left to right) and the 0.3" FWHM HST WFC3 , stellar mass (as inferred from SED fitting, 0.75" FWHM) and mass distribution of the best exponential fit model . The middle row shows the CO velocity map (left), the model velocity map (center), and the data minus model residual map, with superpose white contours of the 0.3" FWHM HST WFC3 map. The bottom row shows the same overlays for the dispersion maps. The black cross denotes the position of the H-band peak that is~0.4"(±0.1") ESE of the stellar mass and star formation rate peaks. Numbers in the individual panels give minimum and maximum values of the color maps, as well as rms and average values of velocity residuals and velocity dispersion residuals.

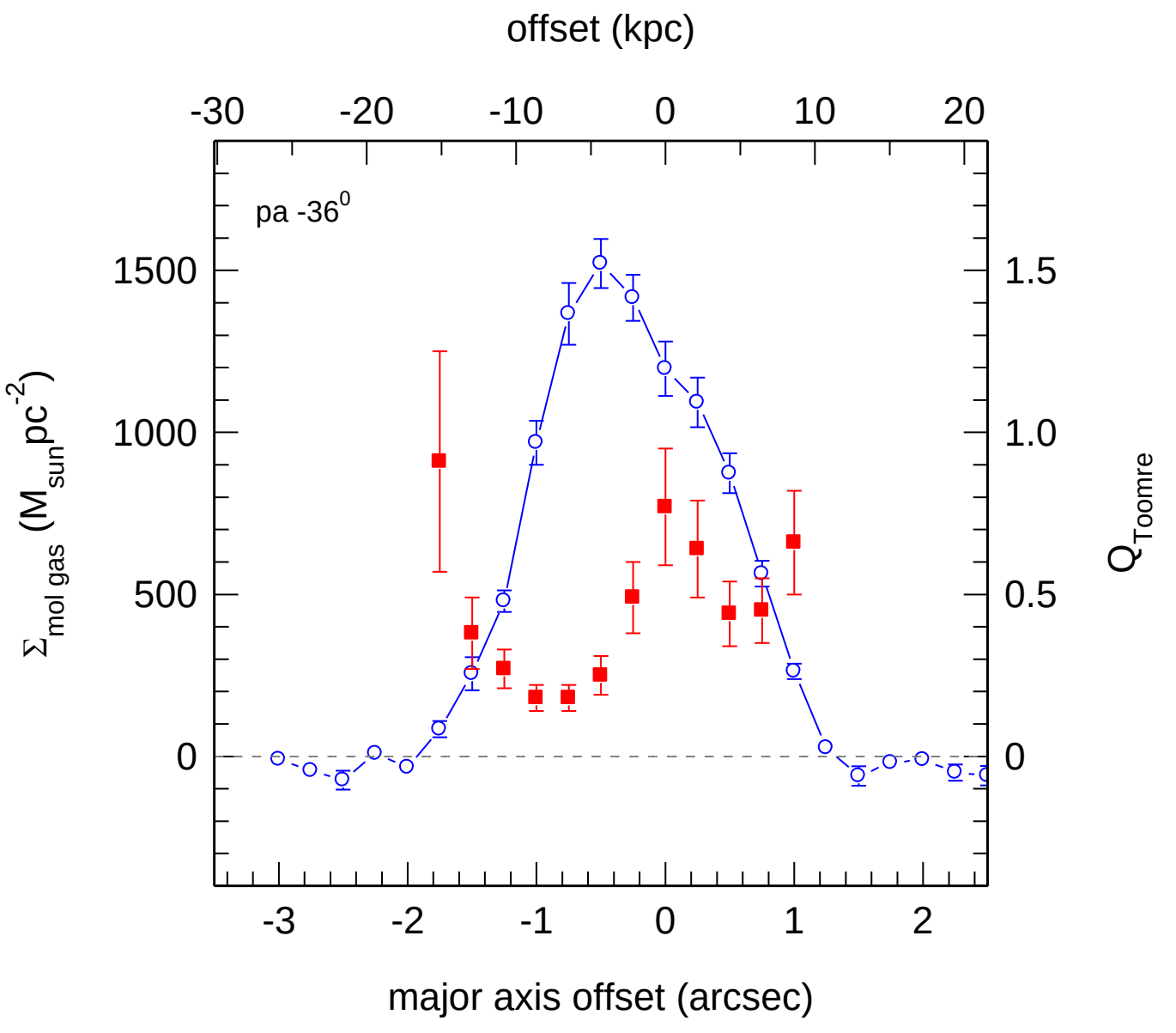

Figure 5. Molecular surface density (open blue circles) and Toomre Q-parameter (filled red squares, $Q_{Toomre} = \frac{\kappa\sigma_0}{\pi G \Sigma_{molgas}}$) in a 0.75" software slit along the major axis of the galaxy. Here κ is the epicyclic frequency, $\kappa^2 = R \times d((v/R)^2)/dR + 4\times(v/R)^2$, $v$ is the rotation velocity at radius $R$, $\sigma_0$ the average local velocity dispersion (after removal of rotation and correction for instrumental resolution) and $\Sigma_{molgas}$ is the molecular gas surface density.

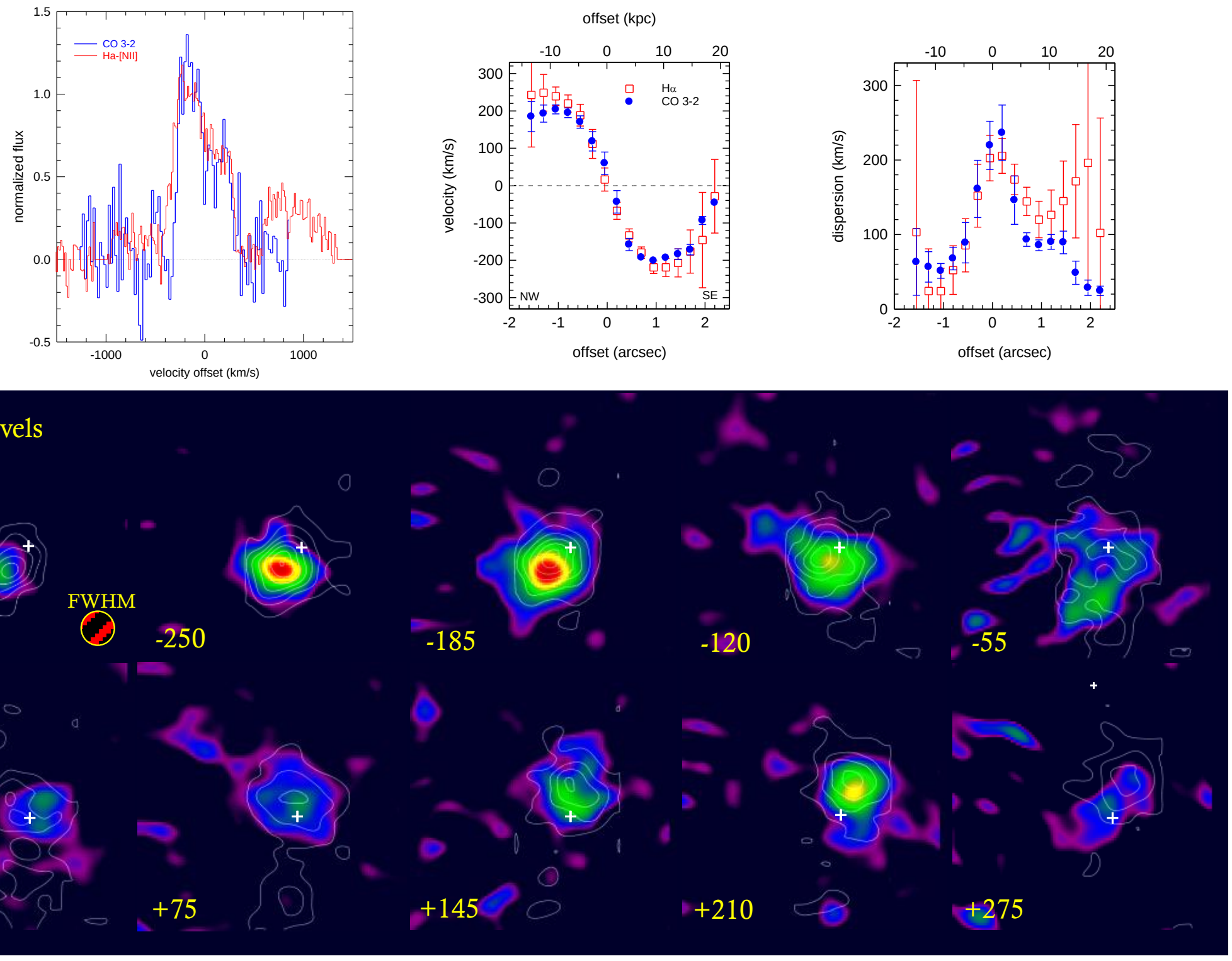

Figure 6. Comparison of CO and Hα data at the same resolution of FWHM 0.75”. The top left panel is a comparison of the integrated, normalized CO (blue) and Hα (red) spectral profiles, with an instrumental resolution of FWHM 22 km/s (CO) and 100 km/s (Hα). The bumps at +940 and -670 km/s in the Hα spectrum are due to [NII]. Top central and right panels: Velocity dispersion and velocity (and ±1σ error) in a software slit of 1” width, from Gaussian fitting along the major kinematic axis (position angle $36^0$ west of north) in CO (blue) and Hα (red). Central and bottom rows: Velocity channel maps (width ~65 km/s) in Hα (contours) and CO (color), sampled onto a 0.06”/pixel grid.

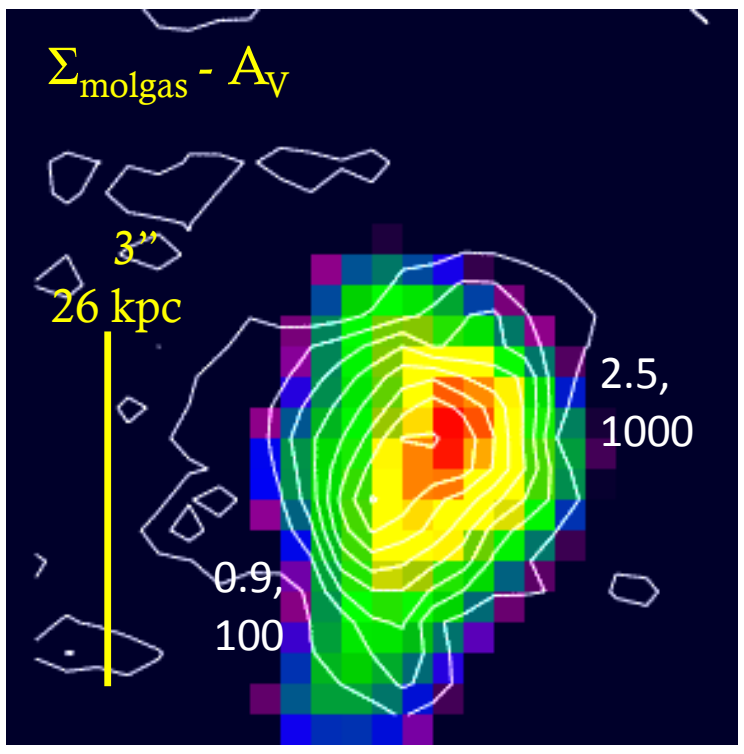

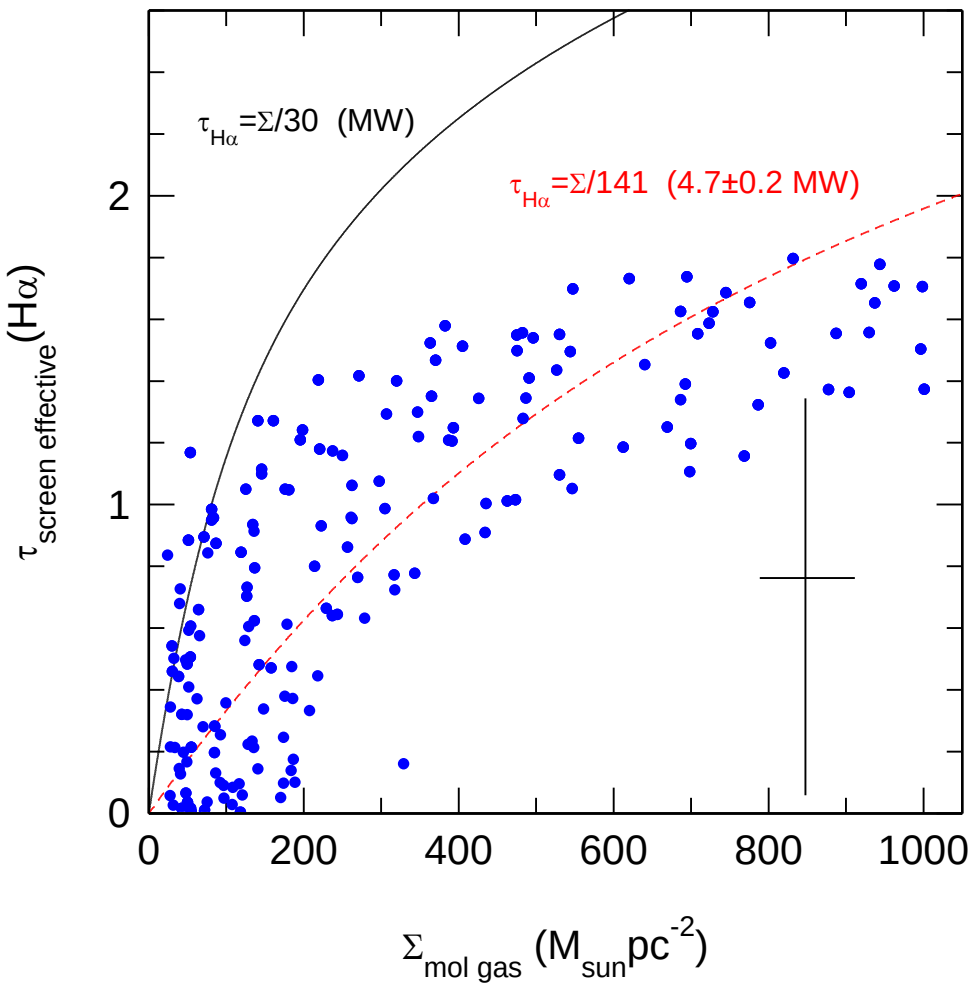

Figure 7: Comparison of effective screen extinction, as inferred from SED analysis of the 4-band HST images, and the CO molecular gas mass surface density $\Sigma_{\text{mol gas}}$. Left panel: comparisons of the two dimensional distributions (CO in white contours, A(V) in color) at the same FWHM resolution of ~0.75", and sampled on a 0.25" grid. Right panel: pixel by pixel correlation between molecular column density (horizontal scale) and effective screen optical depth at Hα (τ(Hα)=0.73 × A(V), Calzetti et al. 2000), with a typical uncertainty marked as a black cross. The black and dotted red curves are models for a homogenous mixture of gas, dust and stars, where $\tau_{effective\ screen}(H\alpha) = \ln\left((\Sigma_{molgas}/30b)\Big/\left\{1-\exp(-(\Sigma_{molgas}/30b))\right\}\right)$, where b is the ratio of total molecular gas column to Hα optical depth in units of the diffuse Galactic ISM value $(\Sigma/\tau(H\alpha))_{\text{MW ISM}}=30\ M_\odot pc^{-2}$, or $(N(H)/A(V))_{\text{MW ISM}} = 2\times10^{21}\ cm^{-2}/mag$ (Bohlin

et al. 1978, see also Wuyts et al. 2011a, Nordon et al. 2013). The best fitting curve has b=4.7±0.3, in good agreement with the findings of Nordon et al. (2013) for z~1-3 main-sequence galaxies observed in the UV, Herschel PACS and CO.

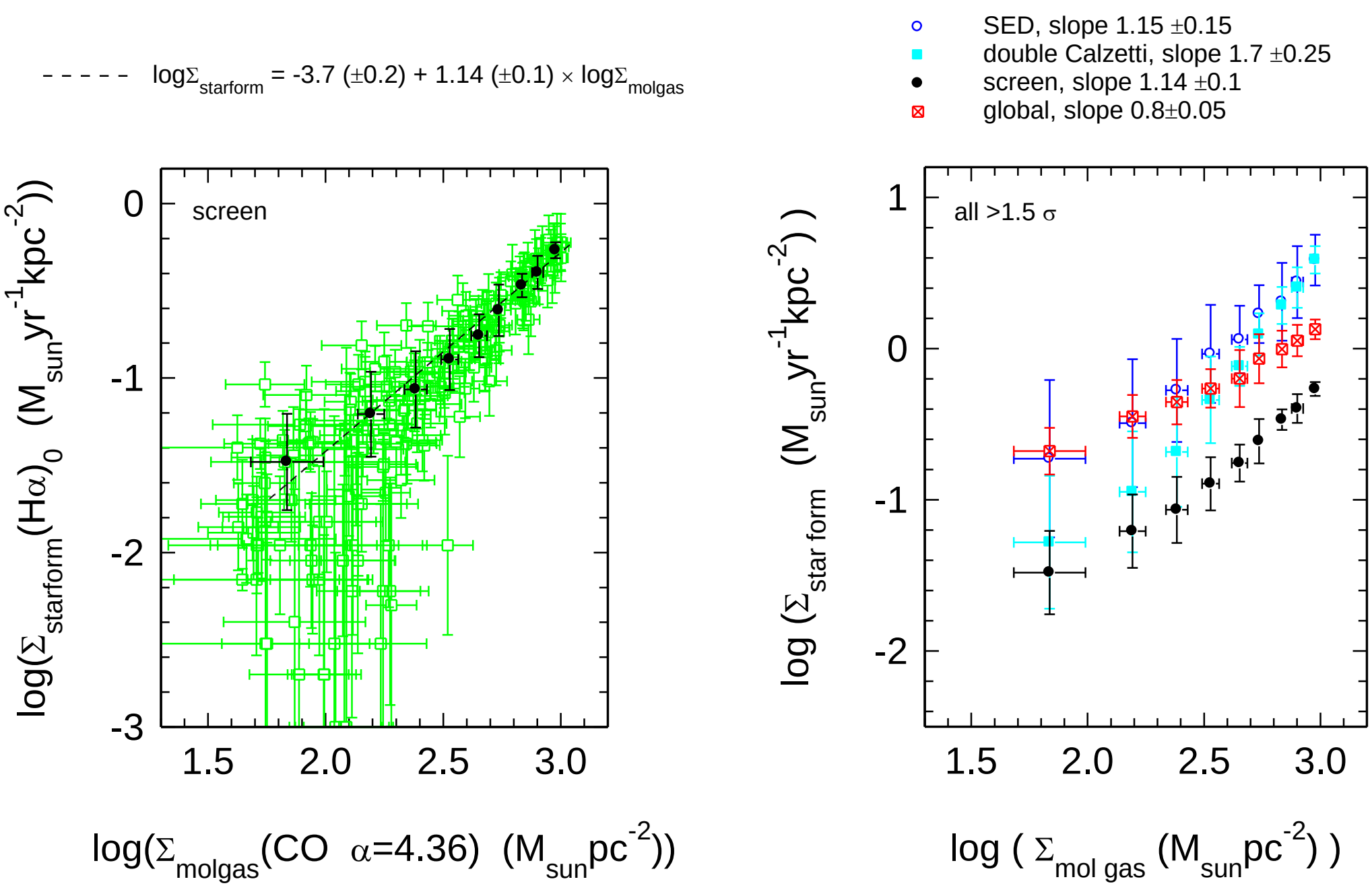

Figure 8. Spatially resolved molecular KS-relation in EGS13011166. Left panel: green circles (and ±1σ uncertainties) denote CO-inferred molecular gas surface densities (horizontal) and the mixed-extinction/effective single screen corrected Hα-star formation rates (with a 'single Calzetti' screen, Calzetti et al. 2000) for each 0.25"x0.25" pixel in the map. Black filled circles and crosses denote the weighted binned medians of the individual pixel values and their dispersions. The dotted black line is the best fit weighted linear regression to the green data, which yields a slope N=1.14±0.1. Right panel: binned weighted medians of the KS-relations for other extinction correction methods: filled black circles denote the mixed case for an effective single Calzetti screen (N=1.14±0.1, the same as in the left panel); crossed red squares denote the global extinction correction method (N=0.8±0.05); the filled cyan squares mark a ‘double Calzetti’ screen correction with an extra factor 2.27 for the nebular over the stellar extinction (N=1.7±0.25); and open blue circles show the correlation for the direct broad-band SED modeling (without using the Hα data) of the star formation rate surface density (N=1.15±0.15).

# Table 1: properties of EGS 13011166

| | |
|---|---|
| coordinates | RA=$14^h19^m45.15^s$ Dec=$52^0 52'27.6''$ |
| redshift | 1.53068(5) |
| $R_{1/2}$ (kpc) (8.62 kpc/arcsecond) | 6.3±1 (UV/optical), 6.3±1 (CO) |
| b/a, i | 0.5 (+0.15,-0.1), $60\pm15^0$ |
| $n_{Sersic}$ | 1±0.5 |
| $v_{rot}$, $v_c$ (km/s) | 307±50, 320±54 at peak |
| $\sigma_0$, $v_{rot}/\sigma_0$ | 52±10, 6.1±2 |
| F(Hα) ($10^{-16}$ erg/s/cm$^2$) | 4.5±1 |
| $F_{co\ 3\text{-}2}$ (Jy km/s) | 2.1±0.2 |
| $F_{[NII]\ 6585}/F_{H\alpha}$, $F_{[SII]\ 6718}/F_{H\alpha}$<br>12+log{O/H}$_{PP04}$, logU<br>12+ log{O/H}$_{D02}$ | 0.37±0.01, 0.25±0.04<br>8.65, -3.5±0.3<br>8.77 |
| $M_*$, $M_{mol\ gas}$, $M_{dyn}$<br>($10^{11}$ $M_\odot$) | 1.2±0.4, 2.6(±0.8)×$\alpha_{4.36}$, 3.0 (±0.6)×$(0.81/\sin(i))^2$ |
| $SFR_{UV}$, $SFR_{SED\ 0}$, $SFR_{FIR}$<br>$SFR_{H\alpha\ 0}$, $SFR_{H\alpha\ 00}$ ($M_\odot$ yr$^{-1}$)<br>$SFR_{tot}$= $SFR_{UV}$+$SFR_{FIR}$ | 23±7, 150±30, 352±80<br>114±25, 560±130<br>375±80 |
| $<\Sigma_{mol\ gas}>$ ($M_\odot$ pc$^{-2}$), $<Q_{Toomre}>$<br>$<\Sigma_{star\ form}>$ ($M_\odot$ yr$^{-1}$ kpc$^{-2}$) | $10^3$, 0.45±0.25<br>1.5 |
| $\tau_{depl}$ = $M_{mol\ gas}/SFR_{tot}$ (Gyr) | 0.7 (galaxy integrated) |
| $\eta_{out}$=$dM_{out}/dt/SFR_{H\alpha\ 00}$ | 1.5 (±1) × ($v_{out}$/600 km/s) × $(R_{1/2}/6.3\ \text{kpc})^{-1}$<br>× $(n_e/50\ \text{cm}^{-3})^{-1}$ |